\documentclass[a4paper,11pt]{article}
\usepackage[T2A]{fontenc}
\usepackage[cp1251]{inputenc}
\usepackage{amsmath,amsthm,amssymb}
\begin{document}
\title{Direct measurements of spin--dependent and coherent effects\\
in conductance of a ferromagnet/superconductor system}
\author{Yu. N. Chiang, O. G. Shevchenko, and R. N. Kolenov\\
\emph{B. I. Verkin Institute for Low Temperature Physics and Engineering},\\
\emph{National Academy of Sciences of Ukraine}\\
\emph{47 Lenin Ave, Kharkov 61103, Ukraine}\\
E-mail: chiang@ilt.kharkov.ua}
\date{}
\maketitle
\renewcommand{\abstractname}{}
\begin{abstract}
In the systems ferromagnet/superconductor ([Fe, Ni]/In), the temperature dependent
transport has been investigated within the temperature range including
superconducting  transition temperature for Indium. It has been found that when
Indium becomes superconducting, the system resistance acquires two positive
additions. The first, numerically equal to the resistance of a ferromagnet part, of
order of a spin--polarized region in length, corresponds to the manifestation of
"spin accumulation" effect. The second agrees in magnitude with an interference
reduction in the ferromagnet conductance over the coherence length for singlet
"Andreev pairs" which is established by the exchange field in the ferromagnet. From
the experimental data, the degree of current spin polarization, the coherence length
in the exchange field, and the lower limit of spin--relaxation length in Fe and Ni
have been estimated.
\end{abstract}
The possibility for spin characteristics of conduction electrons to reveal in
metal transport is widely studied both theoretically and experimentally which
fact indicates that the problems in description and identification of
experimental data in this field is still far from complete. For example, on
measuring transport properties of mesoscopic heterosystems
ferromagnet/superconductor (F/S), an intriguing suggestion has been made [1--3]
that in ferromagnets, a long--range proximity effect for Andreev excitations
with the energies $\varepsilon \sim T < \Delta$ ($\Delta$ is the order
parameter) may exist over the length scale exceeding conventional estimates.
Later analysis [4] of the experiments pointed to some special features
concerning, in particular, the necessity to properly account for the current
distribution in planar mesoscopic interfaces. Besides, the properties of such
interfaces in the form of sandwiches are closely related to the particular
technologies of their preparing. An uncertainty is thus introduced into the
values of the transmission coefficient and potential barrier height at the
interfaces, even for the nanostructures prepared by the technologies of similar
type [5]. Moreover, since the potential difference is usually measured with the
interface included (see, for example, [2, 3]), this circumstance may seriously
interfere with interpreting the effects inherent for a ferromagnet.

As an illustration, the comparative temperature dependences of the potential
difference normalized to the current, {\it U/I}, are plotted in Fig. 1 for the
N/S system normal--metal/superconductor (Cu/Sn) in cases when the probes do or
do not enclose the interface [6]. Curve {\it 1} measured beyond the interface
accounts for the fundamental proposition in theory [7, 8] that at inserting
Andreev reflection, the interference of electrons ({\it e}) and Andreev holes
({\it h}) along the {\it e--h} coherence length, $\xi_{T}$, measured from
N/S interface leads to the increase in elastic scattering cross--section, i. e.
to decreasing (not increasing, as in case the scattering is not considered
[9--11]) the conductance of a normal metal. At the same time, total potential
difference in the whole circuit, with the interface included (most popular
experimental arrangement), after the superconducting transition may bear the
opposite character, not related to the effects in the conductivity of a normal
metal (curve {\it 2}, see also [2]). It is the behavior of this kind that is
often treated as possible manifestation of the long--range proximity effect in
the conductance of a non--superconducting section of the system.

With the aforementioned measurement configuration, the conclusion on the nature
of the processes may also be inadequate when the geometry of the planar
interface is such that a superconductor, as a part of the potential lead,
overlaps a noticeable area $\delta A$ of the non--superconducting film
investigated, and, in doing so, shunts it when turning into the S--state. With
comparatively small thicknesses of the intermediate layer between
superconducting and normal films in the sandwich (the availability of such a
layer is evidenced by the pronounced "non--Sharvin" resistance of the barriers
in mesoscopic interfaces), the inevitable drop in the system resistance due to
the shunting is of order of $\delta R/R \approx \delta A/A$ which value is
often observed [2, 3]. In experiments measuring the "nonlocal" resistance of
the F/S systems [12], the shunting effect may be revealed as well at the
interfaces removed not too far, due to the current spreading into the branches.
The distribution of the electric potential in the branches which size in
mesoscopic samples is comparable to that of the main circuit, is a direct
evidence that the above situation is real. As known, the distribution obeys
the Laplace equation (see, for example, [13]). The solution for the current
near the interface in the branch, {\it L} in length and of width and thickness
identical to those of the main circuit, is the following expression: $j(x)
\approx j(x_{0})(1/4 \sqrt{\pi})(x/x_{0}) \exp[-(x-x_{0})/x_{0}]$. Here,
$x_{0}$ is the beginning of the branch measured from the current injector;
$x_{0} \leq x \leq L$. In mesoscopic samples, the effect should be rather
noticeable. Therefore, in the heterosystems consisted of a ferromagnet and a
superconductor, a few competitive mechanisms may appear after switching into
the F/S regime. The first gives rise to the increase in the potential
difference measured within the range of the coherence length of the
non--superconducting part of the system, due to Andreev interference. The
second diminishes that difference due to the shunting effect (the behavior
of that type was observed in Ref. [12]). The last is the effect at the F/S
interface itself. It is associated with the mismatch between spin--polarized
current in a ferromagnet and spinless current of the Cooper pairs in a singlet
superconductor and is known as spin accumulation [15, 16].

Below we present experimental investigations of transport properties of the
crystalline heterostructures Fe/In and Ni/In, with F/S interfaces prepared
specifically, to obtain high--transparency barriers and, as a result, to
minimize the difference between various F/S interfaces. We have studied the
conductance of the ferromagnet metals with different electron scattering
lengths adjacent to a superconductor, the phenomena at the F/S interface under
Andreev reflection and current polarization, and the shunting effect in the
vicinity of superconducting transition. We use four--terminal technique of
various configurations and different relations between the dimensions of the
ferromagnet conductors and F/S interfaces.

Our main results are as follows: i) We have observed the {\it increase} in the
resistance of F/S interfaces Fe/In and Ni/In as an evidence for spin
accumulation resulted from the peculiarities of Andreev reflection at the F/S
interface under current polarization in a ferromagnet; ii) we have first
revealed the interference contribution from Andreev excitations into the
conductance of a ferromagnet (Ni) within the limits of a coherence length
typical of ferromagnets.

The samples were cut by the spark--erosion method from two bulk ferromagnetic
metals which differ greatly in purity, polycrystalline Fe with Residual
Resistance Ratio $RRR \approx 3$ and
monocrystalline Ni with $RRR \approx 200$. Mean free paths, $l_{el}$, at helium
temperatures were estimated to be approximately 0.01 $\mu$m (the value most
typical of known nanostructures) and 2 $\mu$m for Fe and Ni, respectively.

Fig. 2 presents a schematic view of the sample configurations. The working area
of the samples, with F/S interfaces at the points {\it a} and {\it b} is marked
by dashed lines. When Indium bridge closes the points {\it a} and {\it b} the
working area gains the geometry of an enclosed "Andreev interferometer" which
allows us to study phase--sensitive effects as well (will be presented
elsewhere).

The superconducting In bridge {\it ab} was soldered to the prepared in advance [Fe,
Ni]/In interfaces. Point--like interfaces were fabricated by mechanically destroying
a superficial layer of the ferromagnet metal and simultaneously coating it with
melted Indium of high purity ($RRR \approx 4 \cdot 10^{4}$). Indium was applied onto
a tip of an iron solderer sharpened to the diameter $50 \div 100 \mu$m. The measured
contact resistance of such interfaces does not exceed $1.5 \cdot 10^{-4}\ \Omega$.
To minimize the shunting effect, the ratio between a contact size and sample width
was made down to approximately 0.1. The shunting effect was studied separately on
the Ni sample with purpose designed wide interfaces, see Inset {\it b} to Fig. 4.

The dimensions of the working area elements were as follows. For Fe sample:
width of the ferromagnetic conductors $W_{ac,bd} \approx 1.5$ mm; $W_{cd}
\approx 0.5$ mm; their length $L_{ac,bd} \approx 0.5$ mm; $L_{cd} \approx 0.3$
mm; thickness $t \approx 0.25$ mm. For Ni sample: $W_{ac,bd} \approx 0.5$ mm;
$W_{cd} \approx 0.7$ mm; $L_{ac,bd} \approx 0.5$ mm; $L_{cd} \approx 0.4$ mm
and $t \approx 0.1$ mm. The measuring configuration in each specific case is
shown in the Insets to the Figures.

After curves {\it 1} (Fig. 3, 5) were taken, the resistances of the working
area elements, $R_{ac}(T),\ R_{bd}(T),\ R_{cd}(T)$, were independently measured
by four--terminal technique. In this case, though F/S interfaces were present
their contribution was thereby excluded. Either leads {\it 3, 4} of the sample
or the ends of the opened Indium bridge {\it ab} (see Fig. 2) were used as
potential leads, the current flow configuration remaining fixed.

We achieved a resistance resolution better than $10^{-4}$ due to temperature
and current ($0.1 \div 1$ A) stabilization and using the voltmeter based on a
superconducting modulator accurate within $\delta U \approx 10^{-11}$ V [15]
for measuring potential difference.

Curve {\it 1} in Fig. 3 depicts the temperature dependence of the potential
difference $V_{1}$ and $V_{2}$ at the ends of the {\it cab} part,
$U_{cb}(T)=\vert V_{1}-V_{2} \vert (T)$, normalized to the current in the
branch {\it cabd} formed by two Fe conductors {\it ac} and {\it bd}, two Fe/In
interfaces, and In bridge {\it ab} (see Inset). The current was determined
from the Kirchhoff's laws $I=I_{cabd}(1+I_{cd}/I_{cabd});\ I_{cd}/
I_{cabd}=R_{cabd}/R_{cd}$:
\begin{equation}\label{1}
I_{cabd}(T)=\frac{IR_{cd}-U_{ab}^{\rm In}-2(U_{\rm interface}+\delta U_{ac})}
{R_{\Sigma}}\bigg|_{T};
\end{equation}
$$R_{\Sigma} \vert_{T}=[R_{ac}+R_{bd}+R_{cd}]_{T}; \qquad [U_{\rm interface}+
\delta U_{ac}]_{T}=[U_{cb}-(U_{ab}^{\rm In}+U_{ac})]_{T}.$$
Here, $U_{ab}^{\rm In}$ is the voltage drop measured independently at In bridge,
$U_{\rm interface}$ the potential difference at the interface, $R_{\Sigma}$ the
total resistance of the ferromagnetic part of the contour {\it acdb}, and
$\delta U_{ac}(T)$ a possible addition into the voltage $U_{ac}(T)$ across the
ferromagnet branch {\it ac} which was measured in the configuration excluding
the potential difference at the F/S interface (see above).

As in case with non--magnetic metal in the N/S system Cu/Sn (Fig. 1, curve
{\it 2}), the temperature--dependent resistance of the F/S system Fe/In (Fig.
3, curve {\it 1}) alone does not give an indication of true resistive
contributions from the individual parts that form the F/S system. To separate
those contributions, we should compare curve {\it 1} with the temperature
behavior of the resistance of the same system at the temperatures below
superconducting transition temperature for Indium, $T_{c}^{\rm In}=3.41$ K
(curve {\it 2}) where the resistance of In bridge turns to zero (see Inset
{\it b} to Fig. 3).

Comparing the curves in Fig. 3, $U_{cb}/I_{cabd}(T)$ (curve {\it 1}) and
$R_{ac}(T) \equiv [U_{ac}/I_{cabd}]_{T}$ (curve {\it 2}) at the temperatures
$T \leq T_{c}^{\rm In}$, we conclude that when the interface changes from the
F/N to the F/S state, the resistance of the whole system Fe+F/S interface
increases as opposed to that of the Fe part {\it ac}. Since the latter does not
change significantly (except a hardly visible reduction due to the shunting
effect, see lower panel in Fig. 3) the rise in the former may be attributed to
the effect at the F/S interface. The temperature dependence of the Fe/In
interface resistance, $R_{\rm interface}$, obtained by subtracting the
dependencies $R_{ac}(T)$ and $U_{ab}^{\rm In}/I_{cabd}$ from curve {\it 1} is
shown as curve {\it 2} in Fig. 6.

Fig. 4 displays the resistance $R_{\Sigma}$ of the same part {\it acdb} of
the system Ni/In measured in the presence of wide (curve {\it 1}) or point--like
(curve {\it 2}) F/S interfaces. In both cases, with the measuring
configurations shown (see corresponding Insets), the interface resistance is
excluded as being a part of potential leads. It can be seen that at $T
\leq T_{c}^{\rm In}$, after Andreev reflection turns on, the resistance
$R_{\Sigma}$ increases abruptly by $\approx 1 \cdot 10^{-8}\ \Omega$ in case of
two point contacts ($\delta R/R \approx 0.035$\%) or by $\approx 7 \cdot
10^{-7}\ \Omega$ in case of two wide interfaces ($\delta R/R \approx 2.5$\%).

According to [6--8] the interference contribution from Andreev excitations
scattered at the impurities within a metal layer, of order of coherence length
$\xi$ for Andreev hybrid in thickness, which is measured at a distance {\it L}
from the N/S interface, provided $l_{el} \geq \xi,\ d$ ({\it d} is the
characteristic size of the interface), is given by
\begin{equation}\label{2}
\frac{\delta R}{R}\bigg|_{int}=\frac{\xi}{L}{\overline r}.
\end{equation}
Here, $\overline r$ is the effective probability for each excitation from
an Andreev pair to scatter elastically in the layer $\xi$, regardless of the
number of the pairs, i. e., of the probability of Andreev reflection. The
number of Andreev hybrids is established by the transmission coefficient of an
N/S interface and by the selection rules. The voltage at the interfaces in our
experiments did not exceed 0.15 meV for Fe/In and 3 $\mu$eV for Ni/In, thus, we
can entirely neglect the potential difference at the barrier. Therefore, it
follows from Eq. (2) that, in principle, the resistance of the metal layer,
$L=\xi$ in thickness, may double under Andreev reflection if ${\overline r}=1$.

Consider now ferromagnet metals. On quasiclassical notion, the coherence of
the Andreev pair of excitations in an impure metal is considered to be
destructed when the displacement of their trajectories relative to each other
reaches the value exceeding the size of an impurity (of order of the de Broglie
wave length). The maximum possible distance (collisionless coherence length) at
which this may take place in a ferromagnet exchange field $H_{exch}$ due to the
Larmour curving of {\it e} and {\it h} trajectories, is of order of the Larmour
radius $R_{\rm L}$ [19] and is given by
\begin{equation}\label{3}
\xi_{m} \sim \frac{R_{\rm L}}{2 \pi}=\frac{\hbar v_{\rm F}}{k_{\rm B}T_{exch}},
\qquad l_{el} \geq \xi_{m}
\end{equation}
($k_{\rm B}T_{exch} \equiv \mu H_{exch},\ \mu$ is the Bohr magneton). $\xi_{m}$
is fixed and corresponds to a maximum possible coherence length in a
ferromagnet with elastic scattering length $l_{el} \geq \xi_{m}$. For
$l_{el}<\xi_{m}$ and, hence, low diffusion coefficient {\it D}, the coherence
length is additionally restricted to:
\begin{equation}\label{4}
\xi_{exch}^{D} \sim \sqrt{(1/3)l_{el} \xi_{m}}\ <\ \xi_{m}, \qquad
l_{el}<\xi_{m}.
\end{equation}
For Fe, the Curie temperature $T_{exch}$ is approximately 900 K, and $\xi_{m}
\approx 0.05\ \mu$m. According to the estimates, $\xi_{exch}^{D}$ for our Fe
samples, with $l_{el} \leq 0.01\ \mu$m, is too small in order that the
interference contribution (2) can be detected. In contrast, for Ni, with
$T_{exch} \simeq 600$ K and $l_{el} \sim 1\ \mu$m, the equation (3) is valid,
and theoretical estimate yields $\xi_{m} \sim 0.1\ \mu$m. The coherence length
in our Ni samples is independently evaluated at $\xi \approx 0.1\ \mu$m. We
used Eq. (2) and the experimental data for $\delta R/R$ for the point and wide
interfaces Ni/In. For the wide interface, it was accounted that the number of
Andreev channels should be proportional to the F/S contact area. Therefore,
such an unexpected at first glance manifestation of the coherent effect in a
ferromagnet we observed does not fall outside the scope of {\it conventional}
concepts of the coherence length scale for Andreev excitations in ferromagnetic
metals and has nothing to do with the long--range proximity effect.

The fall on curve {\it 1}, Fig. 4, at $T \approx T_{c}^{\rm In}$ testifies that
when a superconducting potential lead overlaps a part of a normal conductor,
a shunting effect appears due to spreading the current into a superconductor.
The effect may dominate any other effects in conductance of a metal
investigated under the geometry of a three--layered sandwich (in planar
nanostructures) where the resistance of the intermediate layer is, as a rule,
lower than that of a normal metal but higher than that of a superconductor.
That is why we used the technology of preparing the interfaces described above,
which allowed us to get the geometry approaching two--layered one. In this case,
the shunting effect, though visible, is comparable in value to the effect of
interference decrease in conductance of the metal investigated and is at most
twice as much as the latter (see curve {\it 1}, Fig. 4).

Fig. 5 displays the temperature--dependent potential differences for Ni/In
sample with wide interfaces normalized to the current $I_{cabd}$ (see Eq. (1)).
Curve {\it 1}, $U_{ab}^{1}(T)=\vert V_{11}-V_{12} \vert (T)$, was taken
immediately at the Indium bridge while curve {\it 2}, $U_{ab}^{2}=\vert V_{21}
-V_{22} \vert (T)$, was taken at the Ni leads, see Inset to Fig. 5. In both
configurations, the effects beyond the interfaces are eliminated and the
comparison between the two curves allows us to draw a conclusion of the effects
proper to the interface, and their temperature behavior.

In Fig. 6, curve {\it 1}, we present the resistance vs temperature of wide F/S
interfaces in the Ni/In sample (at points {\it a} and {\it b}) as a difference
between curves {\it 1} and {\it 2}, Fig. 5, together with that of Fe/In point
interface (curve {\it 2}, see above). As seen, the resistive behavior of both
wide (Ni/In) and point (Fe/In) interfaces is qualitatively similar,
irrespective of the interface geometry. At $T>T_{c}^{\rm In}$, the interface
resistance varies with temperature as a result of known change in
electron--phonon mean free paths of the metals next to the interface. When
Indium goes to the superconducting state the current component perpendicular to
the plane of an interface disappears. Entering into the superconducting bridge
{\it ab} the current is thus driven to the edges of the interface. "Interface
resistance" at the minimum of curves {\it 1}, {\it 2} can be treated as a
certain minimal value reached at $T=T_{c}^{\rm In}$. It is this value that
should serve as an origin point when calculating any resistive contributions
into the interface resistance if only those may appear when Andreev reflection
turns on. As seen from Fig. 6, these contributions, $\delta R_{F/S}/R_{F/N}$,
are positive for both metals and reach the following values: about 40\% for
Ni/In (curve {\it 1}) and about 20\% for Fe/In (curve {\it 2}).

The above findings we consider as a direct confirmation of spin accumulation at the
interfaces Fe/In and Ni/In. It is clearly demonstrated especially in case of Ni/In
interface where the contribution from the transport effects in a ferromagnetic
branch beyond the interface was almost entirely eliminated. It would appear
reasonable that the value of the resistive jump at the Ni/In interface is entirely
determined by the contribution from small disequilibrium regions close to the
superconducting potential probes where exchange spin splitting takes place. Now we
show that the length of such region, $\lambda_{s}^{*}$, at the Ni/In interface does
not exceed the spin relaxation length, $\lambda_{s}$, in nickel investigated.

In fact, the value for $\delta R_{F/S}/R_{F/N}$ for the sample Fe/In was obtained
in the configuration included the resistance of a {\it long} ferromagnetic
branch, in any case, of no less than a spin--relaxation length in size. It
appeared to be approximately of the same order of magnitude as that for the
sample Ni/In. Therefore, we may conclude that in both our samples, the spatial
scale of the spin--relaxation length $\lambda_{s}$ should be of the same order
of magnitude. As known from theory [15, 16, 4], the change in the resistance
of the F/S interface due to spin--accumulation effect, $\delta R_{F/S}$, is
comparable to the resistance of a ferromagnet part of the length equal to the
spin--flip length:
\begin{equation}\label{5}
\delta R_{F/S}=\frac{\lambda_{s}}{\sigma A}f(P),
\end{equation}
$$f(P)=\frac{P^{2}}{1-P^{2}}; \qquad P=(\sigma_{\uparrow}-\sigma_{\downarrow})
/\sigma; \qquad \sigma=\sigma_{\uparrow}+\sigma_{\downarrow}.$$
Here, {\it P} is the degree of spin polarization; $\sigma,\ \sigma_{\uparrow},\
\sigma_{\downarrow}$, and {\it A} the total, spin--dependent conductivities, and
the cross--section of a ferromagnet, respectively. Using Eq. (5) and geometric
parameters for our samples and taking into account that $P^{Fe} \approx P^{Ni}$
[18] we find
\begin{equation}\label{6}
\frac{\lambda_{s}^{Fe}}{\lambda_{s}^{Ni}}=\frac{\delta R_{Fe/S}}{\delta R_{Ni/S}}
\cdot \frac{l_{el}^{Fe}}{l_{el}^{Ni}} \cdot \frac{A^{Fe}}{A^{Ni}} \approx 2.
\end{equation}
This result confirms that the spatial scales of the spin--flip length, $\lambda
_{s}^{Fe}$ and $\lambda_{s}^{Ni}$, in our samples are comparable. Therefore,
the size of the region which determines the value of the spin--accumulation
effect observed is no more than the spin--relaxation length in each metal. In
this case, according to Eq. (5), it is the values of $\lambda_{s}$ for Fe/In and
$\lambda_{s}^{*}$ for Ni/In that we should use as the length of the conductors
to the resistance $R_{F/N}$ of which we compare the values $\delta R_{F/S}$
obtained in the experiment. It allows us to estimate independently the degree
of conductance spin polarization for Fe and Ni from our experimental data:
\begin{equation}\label{7}
P=\sqrt{(\delta R_{F/S}/R_{F/N})/(1+\delta R_{F/S}/R_{F/N})},
\end{equation}
whence it follows that $P^{Fe} \approx 45$\% and $P^{Ni} \approx 50$\%. The values
coincide practically with those obtained from other experiments [20]. We can roughly
estimate the spin--relaxation length in the metals investigated if by {\it A} in Eq.
(5) is meant the cross--section of the contour through which the current is injected
into the superconducting bridge, i. e., the production of the interface contour
length by the width of the Meissner layer. This yields $\lambda_{s}^{Fe} \sim 900$
\AA\ and $\lambda_{s}^{Ni}>500$ \AA.

In conclusion, we have investigated spin--dependent conductance of the macroscopic
heterosystems ferromagnet (Fe, Ni) / superconductor (In) and obtained further
evidence for the spin accumulation to exist at the F/S interfaces. The effect
results from the peculiarities of Andreev reflection under current polarization in a
ferromagnet. Previously, the experiments on the system Ni/Al of submicron size [21]
have led to the similar conclusion. In addition, we have first proved that the
coherent effects in the conductance of a ferromagnet (nickel) contacted to a
superconductor can be observed within the limits of a coherence length for Andreev
excitations {\it typical} of a ferromagnet, provided the ferromagnet is pure enough.
Our experiments verify that the coherence length in a pure ferromagnet may appear to
be comparable to that length in a non--magnetic metal with shorter elastic
scattering length of electrons. Our results do not confirm the possible existence of
the long--range proximity effect in conventional ferromagnets.

\newpage
\begin{center}
Figure captions
\end{center}

Fig. 1. Temperature dependence of the resistance of the system normal--metal
/superconductor (Cu/Sn) in the measuring configurations beyond the interface
(curve {\it 1}) and including the interface (curve {\it 2}) [6].

Fig. 2. Schematic view of the samples [Fe, Ni]/In. The conductance measurements
were performed inside the working area {\it acdb} enclosed by the dashed line.

Fig. 3. Upper panel: Temperature dependence of the Fe/In system resistance (curve
{\it 1}) measured in the configuration shown in Inset {\it a} and that of the Fe
part {\it ac} (curve {\it 2}) measured independently.

Inset {\it a}: Configuration of measurements. Thin lines schematically depict
Fe conductors. Arrows indicate the current flow path.

Inset {\it b}: Indium bridge resistance vs temperature.

Lower panel: A section of curves {\it 1}, {\it 2} in the vicinity of Indium
superconducting transition on an enlarged scale.

Fig. 4. Resistance of the ferromagnetic Ni contour {\it acdb} in
wide (curve {\it 1}, Inset {\it b}) or point--like (curve {\it 2}, Inset
{\it c}) contact with the superconducting In probes.

Fig. 5. Curve {\it 1}: Temperature--dependent resistance of the Indium bridge
measured between the potential probes $V_{11},\ V_{12}$ beyond the F/S interface.

Curve {\it 2}: Temperature--dependent resistance of the Indium bridge in series
with Ni/In interfaces measured between the probes $V_{21},\ V_{22}$.

Inset: Configuration of measurements. Thin lines schematically depict Ni
conductors. Arrows indicate the current flow path.

Fig. 6. Temperature dependence of the resistance at the interfaces Ni/In
(curve {\it 1}) and Fe/In (curve {\it 2}). For details see text.
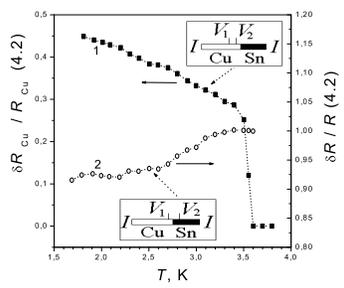
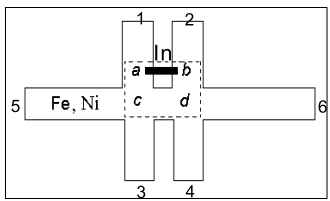
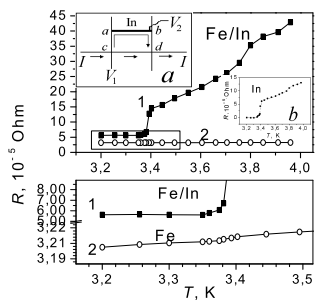
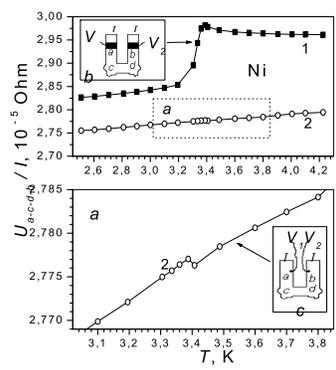
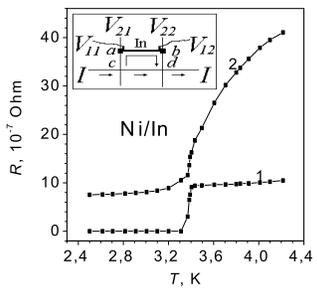
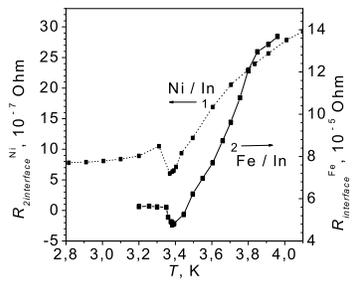
\end{document}